\documentclass{pasj00}
\SetRunningHead{GENJI Programme}{Nagai et al.}

\usepackage{times}
\usepackage{graphicx}
\begin{document}

\title{GENJI Programme: Gamma-ray Emitting Notable AGN Monitoring by Japanese VLBI}
\author{Hiroshi \textsc{Nagai},\altaffilmark{1} \thanks{E-mail: hiroshi.nagai@nao.ac.jp}
Motoki \textsc{Kino},\altaffilmark{1}
 Kotaro \textsc{Niinuma},\altaffilmark{2}
 Kazunori \textsc{Akiyama},\altaffilmark{3,1,8}
 Kazuhiro \textsc{Hada},\altaffilmark{4}
 Shoko \textsc{Koyama},\altaffilmark{3,1}
Monica \textsc{Orienti},\altaffilmark{4,5} 
 Koichiro \textsc{Hiura},\altaffilmark{6}
 Satoko \textsc{Sawada-Satoh},\altaffilmark{7}
Mareki \textsc{Honma},\altaffilmark{1}
 Gabriele \textsc{Giovannini},\altaffilmark{4,5} 
 Marcello \textsc{Giroletti},\altaffilmark{4}
 Katsunori \textsc{Shibata},\altaffilmark{1} \&
 Kazuo \textsc{Sorai} \altaffilmark{6}
}
\altaffiltext{1}{National Astronomical Observatory of Japan, Osawa 2-21-1, Mitaka, Tokyo 181-8588, Japan}
\altaffiltext{2}{Graduate School of Science and Engineering, Yamaguchi University, Yamaguchi 753-8512, Japan}
\altaffiltext{3}{Department of Astronomy, Graduate School of Science, The University of Tokyo, 7-3-1 Hongo, Bunkyo-ku, Tokyo 113-0033, Japan}
\altaffiltext{4}{INAF Istituto di Radioastronomia, via Gobetti 101, 40129, Bologna, Italy} 
\altaffiltext{5}{Dipartimento di Astronomia, Universita' di Bologna, via Ranzani 1, I-40127, Bologna, Italy} 
\altaffiltext{6}{Department of Cosmosciences, Graduate School of Science, Hokkaido University, Kita 10, Nishi 8, Kita-ku, Sapporo, Hokkaido 060-0810, Japan}
\altaffiltext{7}{Mizusawa VLBI Observatory, NAOJ, Hoshigaokacho 2-12, Mizusawa, Oshu, Iwate, 023-0861 Japan}
\altaffiltext{8}{Research Fellow of the Japan Society for the Promotion of Science (JSPS)}


\maketitle

\begin{abstract}
We introduce the GENJI program (Gamma-ray Emitting Notable AGN Monitoring by Japanese VLBI), which is a monitoring program of gamma-ray bright AGNs with the VERA array (VLBI Exploration of Radio Astrometry).  The GENJI programme aims a dense monitoring at 22~GHz towards the $\gamma$-ray emitting active galactic nuclei (AGNs) to investigate the radio time variation of the core and possible ejection of new radio component, motion of jets, and their relation with the emission at other wavelengths especially in $\gamma$-rays.  Currently we are monitoring 8 $\gamma$-ray-emitting notable AGNs (DA~55, 3C~84, M~87, PKS~1510-089, DA~406, NRAO~530, BL~Lac, 3C~454.3) about once every two weeks.  This programme is promising to trace the trend of radio time variation on shorter timescale than conventional VLBI monitoring programme and to provide complimentary data with them (e.g., MOJAVE, Boston University Blazar Project).  In particular, we successfully coordinated quick follow-up observations after the GeV $\gamma$-ray flare in NRAO~530 and 3C~454.3 reported by the {\it Fermi Gamma-ray Space Telescope}.  Here we present the initial results of morphology and light curves for the first 7-month operation. 
\end{abstract}

\section{Introduction}
The non-thermal emission from the relativistic jets emanating from the super massive black holes usually dominates at a wide range of electromagnetic spectrum from radio to $\gamma$-rays.  Some emissions appear up to very high energy range even at GeV and TeV $\gamma$-ray energy ranges.  The location of high energy emission and its production mechanism are long standing problem in AGN jet physics.  With the recent progress by the Large Area Telescope on-board {\it Fermi Gamma-ray Space Telescope} and new generation Cherenkov telescopes such as H.E.S.S., MAGIC, VERITAS, we have a new opportunity to explore the jet physics in connection with $\gamma$-ray emission.  The number of AGN-hosted $\gamma$-ray sources have drastically increased compared to the era of the EGRET/{\it CGRO} \citep{Nolan2012}, including new discoveries of $\gamma$-ray emission from misaligned AGNs (e.g., \cite{Abdo2010a}).  While a number of competing models have been proposed (e.g., \cite{Sikora2009} and references therein), the $\gamma$-ray emission mechanism and the location of emitting region are still disputed.  Quasi-simultaneous multiwavelength study is a key to discriminate such models. 

At radio wavelengths, the high angular resolution provided by Very Long Baseline Interferometry (VLBI) is essential to identify the radio counterpart of the $\gamma$-ray emitting site in AGN jets.  In conventional framework of spectral energy distribution (SED) modeling with one-zone synchrotron-self Compton and/or external Compton model, the blazars should have a subpc-sized emission region or even smaller size \citep{Kubo1998} and the emission region is possibly associated with the region close to the jet base.  Ultimately, VLBI has a potential to identify the exact location of $\gamma$-ray emitting region.  In addition, VLBI is a unique tool to measure the jet motion, which allows us to investigate whether the $\gamma$-rays beamed with the same Lorentz factor as indicated by the VLBI motion. 

The $\gamma$-ray luminosity shows a good correlation with the radio luminosity over more than 4 orders of magnitude \citep{Lister2011}.  This seems to indicate a tight connection between $\gamma$-ray and radio properties in AGN jets.  So far considerable efforts have been expended to do intensive VLBI monitoring in the context of radio-$\gamma$-ray connection.  However, not all the $\gamma$-ray events in blazars and radio galaxies have a clear radio counterpart as in the case of the GeV-TeV $\gamma$-ray flare in 3C 279 \citep{Abdo2010b} and GeV $\gamma$-ray flare in 3C~84 (Nagai et al. 2012).  Another intriguing case in PKS~1510-089 is that some $\gamma$-ray flares seem to be related to changes in radio band (e.g., \cite{Marscher2010a}; \cite{Orienti2011}) while others show no relation (e.g., \cite{DAmmando2009}).  Therefore, the location of radio counterpart of the $\gamma$-ray emission is still an open question and additional validations in more samples are highly awaited.
 
Gamma-ray Emitting Notable AGN Monitoring by Japanese VLBI (GENJI) programme aims dense monitoring of the $\gamma$-ray bright AGNs using the VLBI Exploration Radio Astrometry (VERA).  VERA aims at Galactic maser astrometry, but for technical reason VERA needs to observe a bright calibrator every $\sim80$~min.  We use this calibrator time as the slot for GENJI sources.  Since most of the VERA projects are carried out at 22~GHz, GENJI observation is also basically limited to 22-GHz observation.  But, sometimes we can perform 43-GHz observation depending on the aim of VERA project (e.g., SiO maser observation) \footnote{The VERA has no frequency agility, and therefore we cannot do 22 and 43~GHz observations at the same time.}.  The goal of GENJI program is to study the correlation of time variability between $\gamma$-ray and radio components including the radio core and some other jet components in order to identify a possible radio counterpart of the $\gamma$-ray emission.  We also aim to measure the apparent motion of the jet allowing us to calculate the jet speed.  We focus on the time variation on the timescale of shorter than 1 month and quick follow-up after the $\gamma$-ray flare.  Currently, we are carrying out this monitoring once two-weeks.  Such a dense monitoring is a unique at 22~GHz in the northern hemisphere ({\it cf.} TANAMI in the southern hemisphere: \cite{Ojha2010}) so that we can obtain complementary data with other monitoring projects (e.g., MOJAVE\footnote{http:$\slash\slash$www.physics.purdue.edu$\slash$MOJAVE$\slash$index.html}: \cite{Lister2009}; Boston University Blazar Projects\footnote{http:$\slash\slash$www.bu.edu$\slash$blazars$\slash$research.html}).  

The GENJI observations started in 2010 November.  We have already detected radio flaring associated with a GeV flare in PKS~1510-089 (Orienti et al. 2012).  In addition, we have also successfully coordinated observations right before and after the GeV flare in 3C~454.3 and NRAO~530, and these results are now in preparation.  In this paper, we present a quality of the data and initial results for first 7-month data.    Throughout the paper, we adopt the following cosmological parameters; $H_{\rm 0}=70.2$~km~sec$^{-1}$~Mpc$^{-1}$, $\Omega_{\rm M}= 0.27$, and $\Omega_{\rm \Lambda}=0.73$ \citep{Komatsu2011}. 

\section{Sample}
\begin{figure*}[]
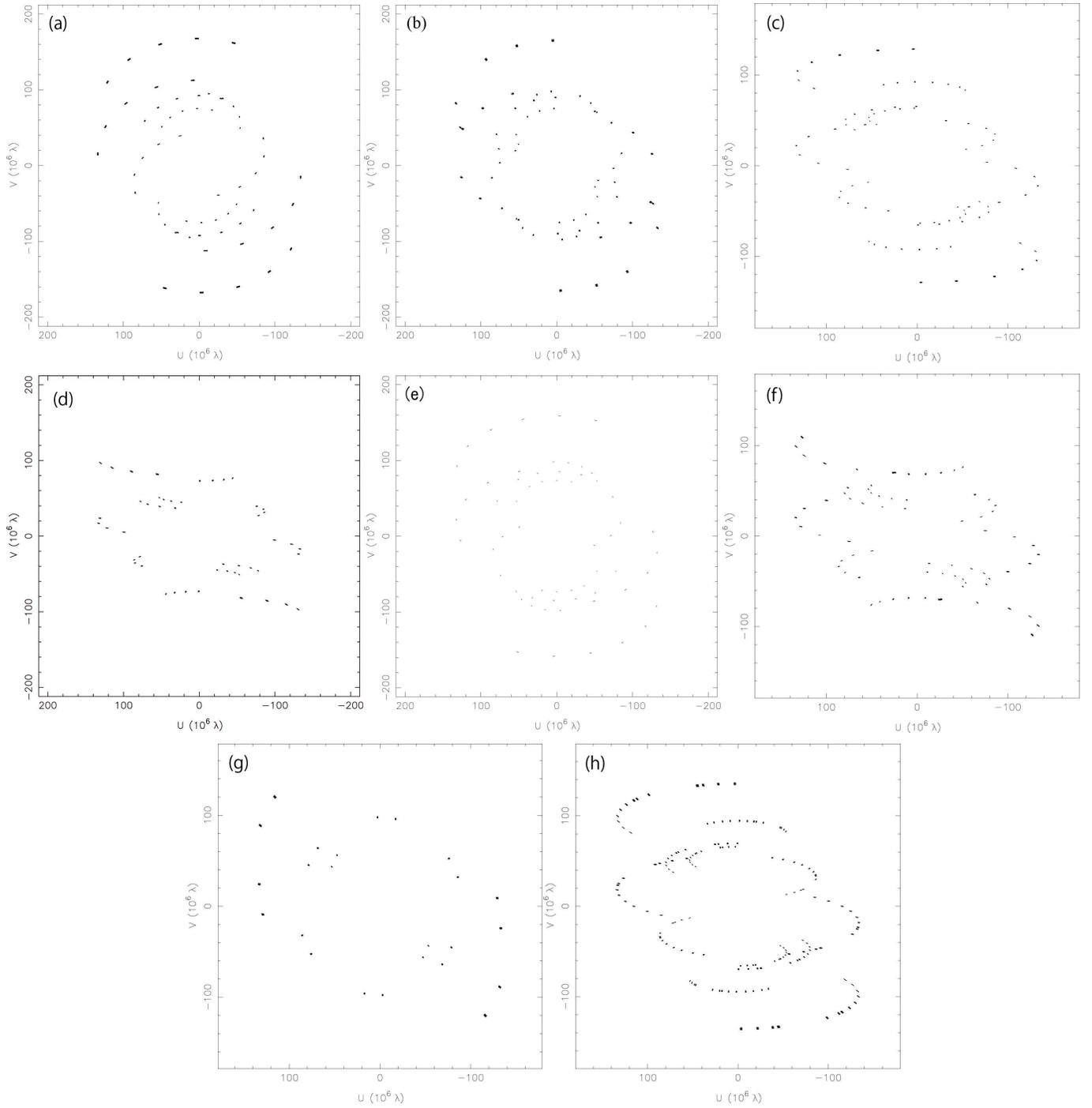
 
\begin{tabular}{cc}
\begin{minipage}{0.33\hsize}
\FigureFile(60mm,40mm){Figure1a.eps}
\end{minipage}
\begin{minipage}{0.33\hsize}
\FigureFile(60mm,40mm){Figure1b.eps}
\end{minipage}
\begin{minipage}{0.33\hsize}
\FigureFile(60mm,60mm){Figure1c.eps}
\end{minipage}\\
\begin{minipage}{0.33\hsize}
\FigureFile(60mm,40mm){Figure1d.eps}
\end{minipage}
\begin{minipage}{0.33\hsize}
\FigureFile(60mm,60mm){Figure1e.eps}
\end{minipage}
\begin{minipage}{0.33\hsize}
\FigureFile(60mm,60mm){Figure1f.eps}
\end{minipage}\\
\begin{minipage}{0.33\hsize}
\FigureFile(60mm,60mm){Figure1g.eps}
\end{minipage}
\begin{minipage}{0.33\hsize}
\FigureFile(60mm,60mm){Figure1h.eps}
\end{minipage}
\end{tabular}
\caption{Examples of $uv$-coverage (a) DA~55, (b) 3C~84, (c) M~87, (d) PKS~1510-089, (e) DA~406, (f) NRAO~530, (g) BL~Lac, (h) 3C~454.3.}
\label{fig:uv}
\end{figure*}

We do not demand statistical completeness of AGN jets, and we rather aim dense sampling for particular sources.  The GENJI sample consist of seven bright blazars and two radio galaxies (3C~84 and M~87).  The sample is selected from the VLBA 2-cm Survey (Kellermann et al. 1998) and MOJAVE programme \citep{Lister2009}, whose expected correlated flux density is brighter than $\sim1$~Jy to fulfill a role of fringe finder or delay calibrator for the VERA project observations, identified as GeV $\gamma$-ray source by {\it Fermi}/LAT.  Because GENJI observations should be carried out within a short time slots not to disrupt total observing time for the VERA project by antenna slew as much as possible, we can basically choose the sources close to the target source of the VERA project on the celestial plane.  We finally selected 8 source listed in Table \ref{tab:sources}.  The plots of typical interferometric $uv$-coverage are shown in Figure \ref{fig:uv}.  OJ~287 has been included in the GENJI sample since 2011 December.

\begin{table*}
 \caption{Sources}\label{tab:sources}
 \begin{center}
   \begin{tabular}{cccccc} \hline\hline
 Source & Alias & z & pc/mas \footnotemark[$*$] & Optical ID. \footnotemark[$\dagger$] & \\  \hline 
J0136+4751 & DA~55  & 0.859 & 7.759 & FSRQ   \\ 
J0319+4130 & 3C~84  & 0.0176 & 0.344 & NLRG/Sy2     \\
J1230+1223 & M~87   & 0.004360 & 0.112 & NLRG   \\
J1512-0905 & PKS~1510-089  & 0.36 &  5.043 & Sy1/HPQ   \\
J1613+3412 & DA~406  & 1.39712 & 8.568 & FSRQ   \\
J1733-1304 & NRAO~530  & 0.902 & 7.879 & FSRQ/LPQ   \\
J2202+4216 & BL~Lac  & 0.0686 &  1.289 & BLLAC  \\ 
J2253+1608 & 3C~454.3  & 0.859 & 7.757 & FSRQ/HPQ  \\ \hline 
\multicolumn{4}{@{}l@{}}{\hbox to 0pt{\parbox{110mm}{\footnotesize 
\footnotemark[$*$] Linear size of 1 milli-arcsec. \\
\footnotemark[$\dagger$] FSRQ: flat spectrum radio quasars, NLRG: narrow line radio galaxies, Sy1: Seyfert type-1, Sy2: Seyfert type-2, HPQ: highly polarized quasars, LPQ: low polarized quasars, BLLAC: BL Lac objects 
}\hss}}
   \end{tabular}
 \end{center} 
\end{table*}

\section{Observation and Data Reduction}\label{sect:obs}
Observations were carried out using the VERA 4 stations.  The monitoring was suspended from 2011 March 11 to 2011 April 19 because of a big earthquake hitting north-east Japan.  From 2011 June to 2011 August, the VERA was not able to be used because of the maintenance.  Most of the sources were observed at 22~GHz.  Sometimes 43-GHz observations were also performed, but we do not include the 43-GHz result in this paper.  On-source time for each observation was typically 30~minutes, spreading over 4-6 scans at different hour angle.  Each source of the sample was observed almost once every two weeks.    

Left hand circular polarization (LHCP) was received and sampled with 2-bit quantization, and filtered using the VERA digital filter unit \citep{Iguchi2005}.  The data were recorded at a rate of 1024~Mbps.  The baseband allocation depends on the observation, but mostly providing a band-width of 256~MHz in which 14 IF-channels per a total of 15 IF-channels of 16~MHz bandwidth were assigned to the GENJI source.  Because of a significant signal loss ($\sim30$\%) due to the analog filter implemented in front of the digital filter, we do not use the data of 12th, 13th and 14th IFs.  Correlation processes were performed with the Mitaka FX correlator \citep{Chikada1991}.

Data reduction was performed using the NRAO Astronomical Image Processing System (AIPS).  At the first step of the data reduction, we flagged the data of 10-channels at both edge of the band in each IF, and then normalized the cross-correlation by the auto-correlation.  A standard {\it a priori} amplitude calibration was performed using the AIPS task APCAL based on the measurements of the system temperature ($T_{\mathrm{sys}}$) by the R-Sky method during the observations and the aperture efficiency provided in the VERA Status Report in 2009.  We did not employ the opacity correction in the task APCAL since the $T_{\mathrm{sys}}$ derived from the R-sky method includes the effect of atmospheric absorption.    
This amplitude calibration provides an accuracy of $\sim10$~\%, according to a number of experience of VERA observations (e.g., \cite{Petrov2012}).  Fringe fitting was done using the AIPS task FRING.  After careful flagging of bad data and averaging 10 sec in time domain and over a passband, we constructed initial source model by the CLEAN algorithm implemented in Difmap software package and did self-calibration by adjusting the visibility phase to minimize the residual error with conserving the closure phase.  After performing a number of iterations with the CLEAN and phase self-calibration and the residual error between the model and observed visibility converged, we switched to both phase and amplitude self-calibration if it works.  
 
For some observations, we combined multiple $uv$-data sets taken within a few days, to improve the $uv$-coverage.  This is done by the AIPS task DBCON after {\it a priori} amplitude and phase calibrations.  After combined the data sets, we performed the imaging as described above.

\subsection{Modeling the Data}\label{sect:[modelfit]}
To quantify the position, size, and flux density of core and jet components, we modeled the source by multiple Gaussian components.  We fitted the visibility data with elliptical Gaussians that gave a good fit to the data as judged by a relative $\chi^{2}$ statistic using the {\it modelfit} task in Difmap.  We sometimes used elliptical Gaussians or point sources if they can produce better fits and consistency across the epochs.  Figure \ref{fig:[modelfit]} shows examples of model-fit image.  Note that these images are essentially similar to total intensity images produced by the CLEAN (Figure \ref{Fig:[DA55]}-\ref{Fig:[3C454.3]}), demonstrating the validity of modeling procedure.  

\section{Flux Density and Positional Accuracies}\label{sect:[accuracy]}
It is important to check the flux density and positional accuracy for the study of time variation.  Among eight sources, we obtained a subset of two or three images observed within a very short timescale ($\sim1$ weeks) for 3C~84, NRAO~530, 3C~454.3, and BL~Lac.  Using these four sources, we checked the consistency of flux density of each component.  In Table \ref{tab:fluxerror} we calculated $\Delta S=\sigma/\bar{S}$ for each source, where $\sigma$ is a standard deviation of flux density among epochs and $\bar{S}$ is the flux density averaged over the epochs shown in Table \ref{tab:summary-of-DA55-parameters}-\ref{tab:summary-of-3C454.3-parameters}.  The resultant $\Delta S$ for the core component was less than 7.5\%, which did not exceed the typical flux calibration error of VERA $\sim10$\% (see \S\ref{sect:obs}). On the other hand, $\Delta S$ of the other jet components were typically 20\% probably because of the lack of sampling on the {\it uv}-plane.  Thus, we adopt 10\% for the flux error of the core component and 20\% for that of other jet components.  One exception was that all three components (C1, C2, and C3) of 3C~84 show very good agreements in flux density among three epochs ($\Delta S\lesssim6$\%).  We adopt the flux error of 10\% for all components of 3C~84.  

We also checked the consistency of the component position with reference to the core, and found that the positional difference between epochs was typically $\sim0.1$~mas.  However, the result of position for the component C2 of PKS1510-089 was quite variable.  This is probably owing to the existence of multiple sub-components in C2 (see \S \ref{sect:PKS1510}).  In this paper, we assume that typical positional accuracy is about 0.1~mas, but more details will be elaborated in the forthcoming paper. 
\begin{table}
 \caption{Flux error estimation}\label{tab:fluxerror}
 \begin{center}
   \begin{tabular}{ccc} \hline\hline
Source & Component & $\Delta S$\footnotemark[$*$] \\ \hline
3C~84\footnotemark[$\dagger$] & C1 & 0.011	\\
 & C2 & 0.062	\\
@& C3 & 0.057 \\ 
PKS1510-089\footnotemark[$\ddagger$] & C1 & 0.018	\\
& C2 & 0.215 \\ 
NRAO~530\footnotemark[$\S$] & C1+C2 & 0.012 \\
& C3 & 0.208 \\
BL~Lac\footnotemark[$\#$] & C1 & 0.075 \\	
& C2 & 0.18 \\ 
3C~454.3\footnotemark[$**$] & C1+C2 & 0.031 \\ \hline
\multicolumn{3}{@{}l@{}}{\hbox to 0pt{\parbox{70mm}{\footnotesize
\footnotemark[$*$] $\Delta S=\sigma/\bar{S}$, where $\sigma$ is a standard deviation of flux density among epochs and $\bar{S}$ is the 
flux density averaged over the epochs.
\footnotemark[$\dagger$] $\Delta S$ is calculated from the data on 2010/Nov/28, 
2010/Nov/29, and 2010/Dec/04.
\footnotemark[$\ddagger$] $\Delta S$ is calculated from the data on 2010/Nov/29 and 2010/Dec/04.
\footnotemark[$\S$] $\Delta S$ is calculated from the data on 2010/Nov/11, 2010/Nov/14 and 2010/Nov/16.
\footnotemark[$\#$] $\Delta S$ is calculated from the data on 201b/Feb/8, 2011/Feb/12 and 2011/Feb/13. 
\footnotemark[$**$] $\Delta S$ is calculated from the data on 2010/Nov/10 and 2010/Nov/12.
}\hss}}
\end{tabular}
\end{center}
\end{table}

\section{Results} 
\subsection{Images}
In Figures \ref{Fig:[DA55]}-\ref{Fig:[3C454.3]}, we show three images in adjacent observing periods for each source.  We convolved with an identical beam which was the largest beam among three epochs for each source.  For each image, we provide the beam size, contour levels, and peak intensity in the caption.  Image noises are generally between 10 and 30~mJy~beam$^{-1}$.  The noise of 3C~454.3 image was 76~mJy~beam$^{-1}$ since the image dynamic range was limited by very bright (more than 20~Jy) core.  Physical parameters of Gaussian models for these three epochs are summarized in Tables \ref{tab:summary-of-DA55-parameters}-\ref{tab:summary-of-3C454.3-parameters}.  All sources exhibited a bright core and a few additional components in the jet.  Because of the lack of short baselines, we only detected the structure within $\sim3$~mas from the core, but missed extended structures.  

\subsection{Light curves}
In Figure \ref{fig:[lightcurve]}, we plot the light curve of our GENJI programme for the first 7-months and the MOJAVE light curve for the same periods.  For GENJI data, we plot the light curve of each component described in the section \ref{sect:[modelfit]} in addition to that of total flux density.  For MOJAVE data, we plot the light curve of total flux density.  The flux error due to the model-fit error is negligible as compared to the flux calibration error.  We assumed that the flux error of MOJAVE is 5\%.  It is obvious from Figure \ref{fig:[lightcurve]} that the sampling interval of GENJI is much denser than that of MOJAVE.

\section{Consistency Check with MOJAVE on Flux Density}
In this section, we compare our GENJI light curve with the MOJAVE light curve.  MOJAVE provides the flux density every few months and the observing frequency 15~GHz is nearly same with GENJI observing frequency 22~GHz.  Therefore, the MOJAVE data can be a check for the fidelity of GENJI data.  

In particular for PKS1510-089, there are two MOJAVE data which were observed in nearly same periods with GENJI.  The flux densities from GENJI are 1.92~Jy on 2010 November 29 and 2.31~Jy on 2011 February 26, while those from MOJAVE are 1.69~Jy on 2010 November 29 and 2.16~Jy on 2011 February 27.  The flux densities of GENJI are 10-20\% higher than those of MOJAVE in both epochs.  Since the most of flux density originates in the core and the core has a slightly inverted spectrum with a spectral index ($\alpha$) of 0.2 \citep{Sokolovsky2010}, these differences in flux density between GENJI and MOJAVE are reasonable.  

Another good example is BL~Lac.  Although we do not have good overlaps in observing periods between MOJAVE and GENJI, the trend of light curve for the 7-months is very similar between them.  The GENJI flux density is always smaller than the MOAJVE flux density by $\sim1$~Jy, but this difference can be explained by the contribution from the extended jet.  A continuously jet extending up to more than 5~mas from the core is seen on the MOJAVE images \citep{Lister2009}.  This extending jet cannot be time variable as compared to the core.  Therefore, the similar trend between GENJI and MOJAVE demonstrates the repeatability of the flux measurement by GENJI.  

Although there are not so many overlaps, similar trend in light curve is apparent in NRAO~530 and 3C~454.3.

In summary, our GENJI programme is promising to provide dense monitoring data for flux measurements.

\section{Notes on Individual Sources}	
In this section, we describe the morphology and light curve for each source after a brief review of background information and previous observations.
\subsection{J0136+4751 (DA~55)}
DA~55 is classified as Flat Spectrum Radio Quasar (FSRQ) at z=0.859 \citep{Healey2008}.  This source is listed in the {\it Fermi}/LAT bright $\gamma-$ray source (0FGL~J0137.1+4751) though it was not detected by EGRET.  The $\gamma$-ray flux density of DA~55 in 2009 mid-March reached about three times the averaged flux density in the first year light curve \citep{Abdo2010c}. The pc-scale jet feature is one-sided in the VLBA 15 GHz total intensity image \citep{Lister2009}.  The jet moves toward northwest direction. 

We detected the jet feature in the position angle $\sim-45^{\circ}$ within $\sim2$~mas from the core (C1) (Figure \ref{fig:[modelfit]}(a) and Figure \ref{Fig:[DA55]}).  The jet can be fitted by a single Gaussian component C2.  Until 2010, a bright jet component (a few hundreds of mJy) located at $\sim$4~mas in the position angle $\sim-40^{\circ}$ from the core can be seen in the MOJAVE images. This outer component is not visible in both the GENJI and the MOJAVE images because of its faintness since GENJI programme has been started. 

We did not obtain many data for this source so far.  No significant change in flux density was detected.  
	
\subsection{J0319+4130 (3C~84)}
The bright radio source 3C~84 (z=0.0176: \cite{Petrosian2007}) shows a two-sided radio jets/lobes on the north and south side of the core in the central 5~pc \citep{Walker2000, Asada2006}.  In mid-1980s, the radio flux density became exceptionally bright, more than 60 Jy at centimeter wavelengths (e.g., \cite{O'Dea1984}), and then subsequently decreased such that, by the early 2000s, the radio flux density decreased to $\sim10$~Jy.  Since 2005, the radio flux density has started to increase in 2005 \citep{Abdo2009a}.  This source was became to be known as a $\gamma$-ray source after the four-months operation of Fermi \citep{Abdo2009a}, and 3C 84 is the typical example of the sources exhibiting clear time variation in the GeV band among the misaligned $\gamma$-ray AGNs (\cite{Kataoka2010}; \cite{BrownAdams2011}). MAGIC Cherenkov telescope also deteced very-high-energy $\gamma$-ray emission with a spectral cut-off around 100~GeV \citep{Aleksic2012}.

The central 1-pc structure consists of three components C1, C2, and C3 (Figure \ref{fig:[modelfit]}(b)).  The images at same frequency with similar resolution in earlier epochs were presented by \citet{Nagai2010}. 
In Figure \ref{Fig:[3C84]}, the emission slightly extending to the south of C3 can be seen.  This extending emission became more visible in later epochs.  An additional minor component can be fitted at $\sim2$-mas south from the C3 in the last few epochs after 2011 April (Figure \ref{fig:[3C84_2011April21]}).

While there are only two data by MOJAVE, GENJI provides the light curve in more detail (Figure \ref{fig:[lightcurve]}(b)).  There is a difference in flux density between MOJAVE and GENJI by a few Jy, but this difference can be explained by the contribution from the extended jet that is not recovered by VERA observations since more extended emission is detected in MOJAVE images \citep{Lister2009}.  While the flux density increased steadily at 22~GHz and 43~GHz before 2009 (\cite{Nagai2010}; \cite{Suzuki2012}), the light curve presented here does not show flux density change within a level of 10\%.  The light curve at 37~GHz in the same periods were presented in \citep{Nagai2012}, and the 37-GHz flux density does not change very much as well.

\subsection{J1230+1223 (M~87)}
M~87 is a giant radio galaxy at the center of the Virgo cluster with a well-known one-sided jet. Its proximity (z=0.00436: \cite{RinesGeller2008}) provides a linear resolution of
 1~mas = 0.08~pc, corresponding to $\sim$140$R_{\rm s}$ for a mass of the central black hole $M_{\rm BH} \sim 6\times 10^9M_{\odot}$ \citep{GebhardtThomas2009}. This source has been routinely detected at energies of MeV/GeV \citep{Abdo2009c} and TeV (e.g. \cite{Aharonian2006}; \cite{Acciari2009}; \cite{Abramowski2012}). During the flaring
activity at TeV in 2008, simultaneous VLBI observations detected a
strong flux density increase from the radio core \citep{Acciari2009}, whose
location is specified as $\sim$20~$R_{\rm s}$ from the black hole
\citep{Hada2012}. On the other hand, a distinct multi-wavelength
correlation was found for the recent TeV flaring event in April 2010,
where the core was relatively stable in radio bands
\citep{Hada2012}. Note that the M~87 jet has another remarkable
feature called HST-1 located at a deprojected distance of at least 120~pc, although its direct connection with $\gamma$-ray activity is still under hot debate (e.g. \cite{Chang2010}; \cite{Chang2010}; \cite{Giroletti2012}). 

The images in Figure \ref{Fig:[M87]} shows a core and jet feature.  Although we fitted the image using only one Gaussian component in Figure \ref{fig:[modelfit]}, it is sometimes possible to put an additional component in the jet.  However, it highly depends on the {\it uv}-coverage whether we can put this component.  Thus, we present the model-fit results only for the core (C1).   

The flux density of C1 is around 1~Jy and slightly time variable with a level of 10-20\% within the observing periods presented in this paper (Figure \ref{Fig:[M87]}).    

\subsection{J1512-0905 (PKS~1510-089)}\label{sect:PKS1510}
PKS~1510-089 is a FSRQ at z=0.36 \citep{Thompson1990}.  This source was discovered as a $\gamma$-ray source by EGRET
and one of the three FSRQs detected by MAGIC and H.E.S.S. (\cite{Cortina2012}, \cite{Wagner2010}).  Since 2008 the $\gamma$-ray emission of PKS~1510-089 became highly variable, and many rapid and intense flaring episodes have been detected by
{\it AGILE} and {\it Fermi} (e.g. \cite{DAmmando2009}; \cite{DAmmando2011}).
In 2011 July and October, two major $\gamma$-ray flares were detected.  Especially during the latter flare, the source reached its highest $\gamma$-ray flux density, becoming the second brightest AGN ever observed by Fermi.  Orienti et al. (2012) investigate the connection of the historical $\gamma$-ray activity with lower energy bands and GENJI plays an important role on the paper by providing VLBI core light curve as a part of our first result.  Radio flux density shows no apparent correlation with the former $\gamma$-ray flare, however, if the latter flare is related to radio outburst, the correlation indicates that the $\gamma$-ray emission region locates along the jet about 10~pc downstream from the central engine \citep{Orienti2012}.

We detected the radio emission within $\sim3$~mas from the core (Figures \ref{fig:[modelfit]}(d) and \ref{Fig:[PKS1510]}).  The radio emission mostly originates in the core (C1), and faint jet is extending in the position angle $\sim-30^{\circ}$.  The jet can be represented by a single Gaussian component C2 in our GENJI image.  In the later few epochs, an additional component C3 can be fitted in the region between C1 and C2 (Figure \ref{fig:[PKS1510_2011Nov12]}).  C2 and C3 correspond to N1 and N2 of higher resolution image by VLBA (Fig.4 of \cite{Orienti2011}), respectively.  

We do not include the flux density of C3 in Figure \ref{fig:[lightcurve]}(d) since C3 appeared to be visible only in later few epochs.  Total flux density had decreased from 2.22~Jy on 2010 November 1 to 1.58~Jy on 2010 December 14 (Figure \ref{fig:[lightcurve]}(d)).  It became to be bright in 2011 February (2.04~Jy on 2011 February 11 and 2.26~Jy on 2011 February 26), and then subsequently decreased to about 1.5~Jy in 2011 April.  Overall trend is similar to the MOJAVE, but our GENJI light curve provides more detail change in flux density.

\subsection{J1613+3412 (DA~406)}
DA~406, which is classified as FSRQ, is the highest-redshift object in the GENJI source list ($z=1.397$: \cite{Schneider2007}).  One-sided jet is extending to the southern-east direction in the VLBA 15~GHz image \citep{Lister2009}. A bright component is located at about 4.5~mas from the core.  This source was detected with both EGRET (3EG J1614+3424) and {\it Fermi}/LAT (1FGL J1613.5+3411).  The $\gamma$-ray flux density of this source in EGRET-era was clearly brighter than in {\it Fermi}-era (\cite{PinerKingham1997}; \cite{Abdo2010d}). 

Our GENJI images show a bright radio core and one-sided jet extending in the position angle of $\sim170^{\circ}$ (Figures \ref{fig:[modelfit]}(e) and \ref{Fig:[DA406]}).  The central $\sim2$~mas structure can be fitted with two components C1 and C2.  Another component C3 can be fitted in the jet feature at 4.5~mas from the core.  No significant motion of these components can be seen by GENJI 7-month monitor.

Our GENJI light curve shows that the flux densities in the later periods (from 2011 April 22 to 2011 May 25) clearly increased from those in earlier epochs (Figure \ref{fig:[lightcurve]}(e)).  The MOJAVE provides only one data point in these periods, and therefore this change in flux density is not detected.   

\subsection{J1733-1304 (NRAO~530)}
NRAO~530 (also known as PKS~1730-13) is a well-known blazar at z=0.902 \citep{Healey2008} classified as a FSRQ source. NRAO~530 has a strong variability at various wavelength such as radio \citep{Bower1997}, optical \citep{Webb1988}, X-ray \citep{Foschini2006} and the $\gamma$-ray bands \citep{Mukherjee1997, D'Ammando2010}.  In $\gamma$-ray band, NRAO~530 was identified with the EGRET source 2EG~1735-1312 \citep{Thompson1995}. After the launch of {\it Fermi}, it did not appear as a bright source \citep{Abdo2009b} but as a relatively quiescent source 1FGL~J~1733.0-1308 \citep{Abdo2010b}. However, its $\gamma$-ray activity increased at the latter half of 2010 and NRAO~530 flared between October 31 and November 2, 2010 \citep{D'Ammando2010}.  On pc-scale, NRAO~530 has a core-jet structure extending to the north from the core with a wiggled trajectory on a scale of about 30~mas \citep{Lu2011}. 

\begin{figure*}[]
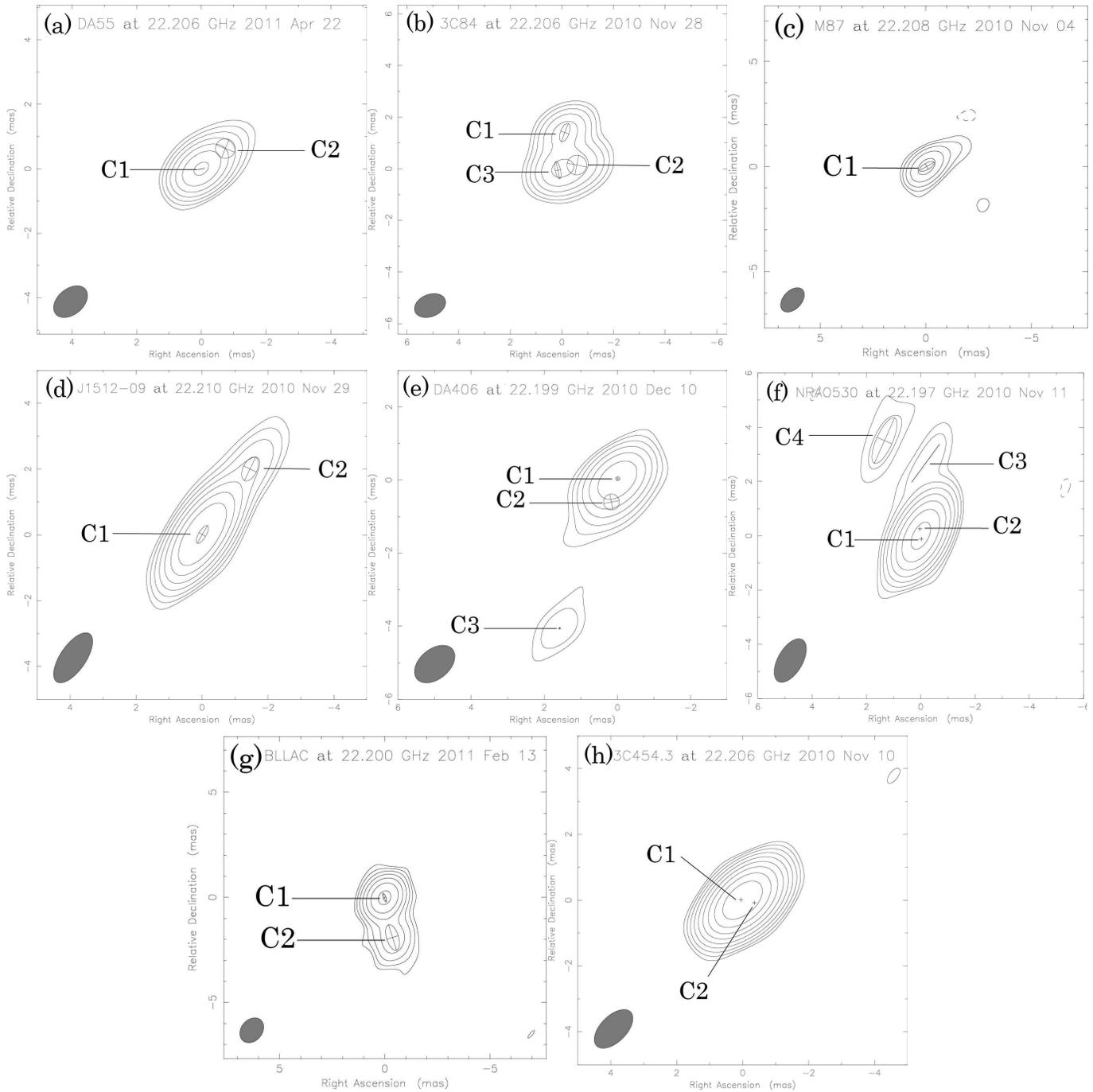
 
\begin{tabular}{ccc}
\begin{minipage}{0.33\hsize}
\FigureFile(60mm,60mm){Figure2a.eps}
\end{minipage}
\begin{minipage}{0.33\hsize}
\FigureFile(60mm,40mm){Figure2b.eps}
\end{minipage}
\begin{minipage}{0.33\hsize}
\FigureFile(60mm,60mm){Figure2c.eps}
\end{minipage}\\
\begin{minipage}{0.33\hsize}
\FigureFile(60mm,40mm){Figure2d.eps}
\end{minipage}
\begin{minipage}{0.33\hsize}
\FigureFile(60mm,60mm){Figure2e.eps}
\end{minipage}
\begin{minipage}{0.33\hsize}
\FigureFile(60mm,60mm){Figure2f.eps}
\end{minipage}\\
\begin{minipage}{0.33\hsize}
\FigureFile(60mm,60mm){Figure2g.eps}
\end{minipage}
\begin{minipage}{0.33\hsize}
\FigureFile(60mm,60mm){Figure2h.eps}
\end{minipage}
\end{tabular}
\caption{Examples of Gaussian model fit of (a) DA~55, (b) 3C~84, (c) M~87, (d) PKS~1510-091, (e) DA~406, (f) NRAO~530, (g) BL~Lac, and (h) 3C~454.3.  Total intensity image is created by CLEAN and model-fit image is produced by restoring the model-fit components. }
\label{fig:[modelfit]}
\end{figure*}

We detected a structure within 5~mas from the core (Figures \ref{fig:[modelfit]}(f) and \ref{Fig:[NRAO530]}). The observed structure is consistent with previous VLBI observations. It mainly consists of two major components C1, C2 and two faint jet components C3 and C4 (Figure \ref{fig:[modelfit]}).  

In Figure \ref{fig:[lightcurve]}, we show the sum of C1 and C2, since the distance between C1 and C2 is very small and the flux density of C1 can be correlated with that of C2.  In some epochs, we could not distinguish C2 from C1 or we could not detect C3 and C4 presumably owing to a lack of $uv$-coverage.  We do not include the light curve of C4 in Figure \ref{fig:[lightcurve]} since the fit is not quite robust.  Our observations started 9 days after the 2010 October $\gamma$-ray flare. Our results show that the core flux density of NRAO~530 had increased from $\sim2.7$~Jy to $\sim4.0$~Jy in the end of 2010. The flared core flux density had been relatively stable until 2011 May and started to decrease in 2011 May.  The change in flux density in this period is difficult to follow only with MOJAVE.  NRAO~530 is one of the most successful examples to show how intensive monitoring provided by GENJI is important to study the connection between $\gamma$-ray flare and the change in radio flux density.

\subsection{J2202+4216 (BL~Lac)}
BL Lacertae (z=0.0686: \cite{Sazonov2007}) is a prototypical blazar which has been well studied across the entire electromagnetic spectrum. This source has been detected by EGRET several times and now routinely detected by $Fermi$ \citep{Abdo2011}. Above 200~GeV, a significant detection was made by MAGIC \citep{Albert2007}. VLBI structure of this source is characterized by the bright core and the extended jet propagating southward (e.g. \cite{Jorstad2005}; \cite{Bach2006}).  Recent multi-wavelength study detected a multiple $\gamma$-ray flaring event which is followed by a emergence of a superluminal knot from the core, together with a continuous rotation of the optical polarization angle \citep{Marscher2008}. These lead to the scenario that the flaring event in this source is produced at large distance (more than 1~pc or 10$^{4}~R_{\rm s}$) downstream of the central engine, although such an interpretation is still controversial yet \citep{Tavecchio2010}.

We detected the bright core (C1) and a jet feature extending to a few milli-arcsec from the core.  The jet feature can be represented as a single Gaussian component C2 (Figure \ref{fig:[modelfit]}(g)).  

The difference in flux density between our GENJI light curve and MOJAVE light curve (see Figure \ref{fig:[lightcurve]}(g)) can be explained by the contribution from the extended jet since a continuous jet extending up to more than 5~mas from the core is seen on the MOJAVE images \citep{Lister2009}.  The flux density decreased from 4.0~Jy on 2010 November 22 to 2.5~Jy on 2011 January 15, and then small rise and decay in flux density is seen around 2011 February to March.  In later epochs, the flux density increased slightly from 2.6~Jy on 2011 March 2 to 3.2~Jy on 2011 May 22. 
\subsection{3C~454.3}
A well-known FSRQ source 3C~454.3 at z= 0.859 \citep{Sazonov2007} is one of the brightest extragalactic radio sources.  After 2000, the radio flux density of 3C~454.3 had been quite time variable and showed increased activity many times.  In particular since 2005, a remarkable flaring activity was seen in this source across the whole electromagnetic spectrum from radio to $\gamma$-rays. During the last four years, this source has exhibited more than one $\gamma$-ray flare every year, becoming the most active $\gamma$-ray blazar in the sky.  The largest flare occurred in November 2010.  In this flare, 3C~454.3 reached a peak flux density more than a factor of 6 higher than the flux density of the brightest steady $\gamma$-ray source, the Vela pulsar, and more than a factor of 3 brighter than its previous super-flare on 2-3 December 2009 \citep{Abdo2011,Vercellone2011}.  On pc-scale, 3C~454.3 has a core-jet structures extending to the west.  Previous VLBI observations showed jet components propagating along different trajectories \citep{Lister2009}.

We detected a structure within 1~mas from the core. It mainly consists of two major components C1, C2 (Figure \ref{fig:[modelfit]}(h)).  

In Figure \ref{Fig:[3C454.3]}, we show the sum of flux densities of C1 and C2, since the distance between C1 and C2 is very small and flux density of C1 can be correlated with C3.  Our observations started just before the largest $\gamma$-ray flare in November 2010. Our results show that the core flux of 3C~454.3 started to increase in 2010 December and peaked around 2011 February.  This is consistent with the idea that the $\gamma$-ray flaring region is located at the upstream of the radio core and the flux density increase in the radio core is the result of the propagation down of the flaring component.  If we assume a jet speed of $\Gamma=10$ ($\Gamma$ is the bulk Lorentz factor), the $\gamma$-ray flaring zone should be $\sim0.05$~pc upstream of the 22-GHz radio core.  We note that this rise and decay in flux density is visible from the GENJI light curve but is not very apparent from the MOJAVE light curve (Figure \ref{fig:[lightcurve]}).

\section{Summary}
We presented the first 7-month results from the GENJI programme.  In this paper, we particularly focused on the fidelity of the image and flux density measurement by comparing nearby epoch data, and we also presented the consistency with the MOJAVE.  We will discuss the detail of light curve and kinematics for individual sources in separate papers.  In particular, we successfully obtained the data right before/after the $\gamma$-ray flares reported in PKS1510-089, NRAO~530, 3C~454.3.  The 22-GHz data from the GENJI are unique and they provide complementary information with other monitoring projects in radio bands.  

\bigskip
We thank the anonymous referee for helpful comments.  We are grateful to the staff of all the VERA stations for their assistance in observations.  This research has made use of the NASA/IPAC Extragalactic Database (NED) which is operated by the Jet Propulsion Laboratory, California Institute of Technology, under contract with the National Aeronautics and Space Administration.  This research has
made use of data from the MOJAVE database that is maintained by the MOJAVE team (Lister et al. 2009). The VLBA is operated by the US National Radio Astronomy Observatory (NRAO), a facility of the National Science Foundation operated under cooperative agreement by Associated Universities, Inc.  Part of this work was done with the contribution of the Italian Ministry of Foreign Affairs and University and Research for the collaboration project between Italy and Japan.  This work is partially supported by Grant-in-Aid for Scientific Research, KAKENHI 24540240 (MK) from Japan Society for the Promotion of Science (JSPS). 


\begin{figure*}[] 
\begin{tabular}{ccc}
\begin{minipage}{0.33\hsize}
\FigureFile(60mm,40mm){Figure3a.eps}
\end{minipage}
\begin{minipage}{0.33\hsize}
\FigureFile(60mm,40mm){Figure3b.eps}
\end{minipage}
\begin{minipage}{0.33\hsize}
\FigureFile(60mm,60mm){Figure3c.eps}
\end{minipage}
\end{tabular}
\caption{ Total intensity images of DA 55. All images are convolved with an identical synthesized beam of (1.2$\times$0.8)~mas at the position angle of $-50$~deg. The contours are plotted at the level of 0.025$\times$(-1, 1, 2, 4,$\dots$, 64)~Jy beam$^{-1}$. The peak intensities of three images are 1.80, 1.83, and 1.84~Jy beam$^{-1}$, respectively.}
\label{Fig:[DA55]}
\end{figure*}

\begin{figure*}[] 
\begin{tabular}{ccc}
\begin{minipage}{0.33\hsize}
\FigureFile(60mm,40mm){Figure4a.eps}
\end{minipage}
\begin{minipage}{0.33\hsize}
\FigureFile(60mm,40mm){Figure4b.eps}
\end{minipage}
\begin{minipage}{0.33\hsize}
\FigureFile(60mm,60mm){Figure4c.eps}
\end{minipage}
\end{tabular}
\caption{Total intensity images of 3C~84.  All images are convolved with an identical synthesized beam of $(1.26\times0.861)$~mas at the position angle of $-67.4^{\circ}$.  The contours are plotted at the level of $18.1\times(-1, 1, 2, 4, \dots, 32)$~mJy~beam$^{-1}$.  The peak intensities of three images are 7.05, 6.79, and 7.27~Jy~beam$^{-1}$, respectively.}
\label{Fig:[3C84]}
\end{figure*}

\begin{figure*}[] 
\begin{tabular}{ccc}
\begin{minipage}{0.33\hsize}
\FigureFile(60mm,40mm){Figure5a.eps}
\end{minipage}
\begin{minipage}{0.33\hsize}
\FigureFile(60mm,40mm){Figure5b.eps}
\end{minipage}
\begin{minipage}{0.33\hsize}
\FigureFile(60mm,60mm){Figure5c.eps}
\end{minipage}
\end{tabular}
\caption{Total intensity images of M~87.  All images are convolved with an identical synthesized beam of $(1.36\times0.87)$~mas at the position angle of $-43^{\circ}$.  The contours are plotted at the level of $18\times(-1, 1, 2, 4, \dots, 32)$~mJy~beam$^{-1}$.  The peak intensities of three images are 0.95, 0.95, and 0.95~Jy~beam$^{-1}$, respectively.}
\label{Fig:[M87]} 
\end{figure*}

\begin{figure*}[] 
\begin{tabular}{ccc}
\begin{minipage}{0.33\hsize}
\FigureFile(60mm,40mm){Figure6a.eps}
\end{minipage}
\begin{minipage}{0.33\hsize}
\FigureFile(60mm,40mm){Figure6b.eps}
\end{minipage}
\begin{minipage}{0.33\hsize}
\FigureFile(60mm,60mm){Figure6c.eps}
\end{minipage}
\end{tabular}
\caption{Total intensity images of J1512-0905 (PKS1510-089).  All images are convolved with an identical synthesized beam of (1.68
$\times$ 0.828)~mas at the position angle of $-30.3^{\circ}$. The contours
are plotted at the level of 14 $\times$ (-1,1,2,4,...,32,64) mJy
beam$^{-1}$. The peak intensity of three images are 1.82, 1.66, and 1.62 Jy
beam$^{-1}$, respectively.}
\label{Fig:[PKS1510]}
\end{figure*}

\begin{figure*}[] 
\begin{tabular}{ccc}
\begin{minipage}{0.33\hsize}
\FigureFile(60mm,40mm){Figure7a.eps}
\end{minipage}
\begin{minipage}{0.33\hsize}
\FigureFile(60mm,40mm){Figure7b.eps}
\end{minipage}
\begin{minipage}{0.33\hsize}
\FigureFile(60mm,60mm){Figure7c.eps}
\end{minipage}
\end{tabular}
\caption{Total intensity images of DA~406.  All images are convolved with an identical synthesized beam of ($1.27\times0.82$)~mas at the position angle of $-50^{\circ}$.  The contours are plotted at the level of $22\times(-1, 1, 2, 4, \dots, 32)$~mJy~beam$^{-1}$.  The peak intensities of three images are 1.26, 1.17, and 1.27~Jy~beam$^{-1}$, respectively.}
\label{Fig:[DA406]}
\end{figure*}

\begin{figure*}[] 
\begin{tabular}{ccc}
\begin{minipage}{0.33\hsize}
\FigureFile(60mm,40mm){Figure8a.eps}
\end{minipage}
\begin{minipage}{0.33\hsize}
\FigureFile(60mm,40mm){Figure8b.eps}
\end{minipage}
\begin{minipage}{0.33\hsize}
\FigureFile(60mm,60mm){Figure8c.eps}
\end{minipage}
\end{tabular}
\caption{Total intensity images of NRAO~530.  All images are convolved with an identical synthesized beam of ($1.78\times0.911$)~mas at the position angle of $-30^{\circ}$.  The contours are plotted at the level of $22\times(-1, 1, 2, 4, \dots, 128)$~mJy~beam$^{-1}$.  The peak intensities of three images are 3.94, 3.98, and 3.89~Jy~beam$^{-1}$, respectively.}
\label{Fig:[NRAO530]}
\end{figure*}

\begin{figure*}[] 
\begin{tabular}{ccc}
\begin{minipage}{0.33\hsize}
\FigureFile(60mm,40mm){Figure9a.eps}
\end{minipage}
\begin{minipage}{0.33\hsize}
\FigureFile(60mm,40mm){Figure9b.eps}
\end{minipage}
\begin{minipage}{0.33\hsize}
\FigureFile(60mm,60mm){Figure9c.eps}
\end{minipage}
\end{tabular}
\caption{Total intensity images of BL~Lac.  All images are convolved with an identical synthesized beam of ($1.3\times1.0$)~mas at the position angle of $-40^{\circ}$.  The contours are plotted at the level of $30\times(-1, 1, 2, 4, \dots,  64)$~mJy~beam$^{-1}$.  The peak intensities of three images are 2.30, 2.08, and 2.30~Jy~beam$^{-1}$, respectively.}
\label{Fig:[BLLAC]}
\end{figure*}

\begin{figure*}[] 
\begin{tabular}{ccc}
\begin{minipage}{0.33\hsize}
\FigureFile(60mm,40mm){Figure10a.eps}
\end{minipage}
\begin{minipage}{0.33\hsize}
\FigureFile(60mm,40mm){Figure10b.eps}
\end{minipage}
\begin{minipage}{0.33\hsize}
\FigureFile(60mm,60mm){Figure10c.eps}
\end{minipage}
\end{tabular}
\caption{Total intensity images of 3C~454.3.  All images are convolved with an identical synthesized beam of ($1.46\times0.797$)~mas at the position angle of $-44.7^{\circ}$.  The contours are plotted at the level of $76.3\times(-1, 1, 2, 4, \dots, 256)$~mJy~beam$^{-1}$.  The peak intensities of three images are 23.5, 22.7, and 23.4~Jy~beam$^{-1}$, respectively.}
\label{Fig:[3C454.3]}
\end{figure*}

\begin{table*}
 \caption{Summary of Gaussian model parameters of DA~55.} \label{tab:summary-of-DA55-parameters}
 \begin{center}
   \begin{tabular}{cccccccc} \hline\hline
 Epoch &  Component & Flux [Jy] & Radius [mas] \footnotemark[$*$] & $\theta$ [deg] \footnotemark[$\dagger$] &  Major [mas] \footnotemark[$\ddagger$] & Axial ratio\footnotemark[$\S$] &  $\phi$ [deg] \footnotemark[$**$]     \\  \hline 
2011/Apr/22 & C1 & 1.803 & 0.012 & 115.4 & 0.212 & 0.000 & -79.4 \\ 
 & C2 & 0.161 & 0.971 & -49.8 & 0.606 & 1.000 & -21.8 \\   
2011/May/06+May/09 & C1 & 1.820 & 0.012 & 123.6 & 0.107 & 0.000 & 71.4 \\
 & C2 & 0.162 & 0.992 & -52.3 & 0.867 & 1.000 & 17.1 \\  
2011/May/19 & C1 & 1.877 & 0.012 & 123.0 & 0.291 & 0.140 & -72.0 \\
& C2 & 0.197 & 1.050 & -48.8 & 0.726 & 1.000 & 0.0 \\ \hline
\multicolumn{4}{@{}l@{}}{\hbox to 0pt{\parbox{170mm}{\footnotesize
\footnotemark[$*$] The radial distance of the component center from the center of the map.
\footnotemark[$\dagger$] The position angle of the center of the component.
\footnotemark[$\ddagger$] The FWHM major axis of the component. 
\footnotemark[$\S$] The ratio of the minor axis to the major axis.
\footnotemark[$**$] The position angle of the major axis.
}\hss}}
  \end{tabular}
 \end{center} 
\end{table*}

\begin{table*}
 \caption{Summary of Gaussian model parameters of 3C~84.} \label{tab:summary-of-3C84-parameters}
 \begin{center} 
   \begin{tabular}{cccccccc} \hline\hline
 Epoch &  Component & Flux [Jy] & Radius [mas] & $\theta$ [deg] &  Major [mas] & Axial ratio &  $\phi$ [deg]     \\  \hline 
2010/Nov/28 & C1 & 5.493  & 1.409  & -4.6  & 0.744  & 0.451  & -20.7 \\
 & C2 & 4.956  & 0.595  & -75.8  & 0.761  & 1.000  & -101.3 \\
 & C3 & 6.895  & 0.165  & 108.9  & 0.680  & 0.308  & 12.7 \\
2010/Nov/29 & C1 & 5.447  & 1.449  & -5.8  & 0.793  & 0.423  & -19.6 \\
 & C2 & 5.020  & 0.602  & -73.1  & 0.757  & 1.000  & -90.0 \\
 & C3 & 6.166  & 0.140  & 99.3  & 0.652  & 0.000  & 14.7 \\
2010/Dec/4 & C1 & 5.568  & 1.361  & -3.4  & 0.768  & 0.401  & -21.3 \\
 & C2 & 5.537  & 0.568  & -77.4  & 0.867  & 1.000  & -104.0 \\
 & C3 & 6.690  & 0.191  & 119.5  & 0.595  & 0.382  & 19.3  \\ \hline
   \end{tabular}
 \end{center} 
\end{table*}

\begin{table*}
 \caption{Summary of Gaussian model parameters of M~87.} \label{tab:summary-of-M87-parameters}
 \begin{center}
   \begin{tabular}{cccccccc} \hline\hline
 Epoch &  Component & Flux [Jy] & Radius [mas] & $\theta$ [deg] &  Major [mas] & Axial ratio &  $\phi$ [deg]     \\  \hline 
 2010/Nov/04 & C1 & 1.131 & 0.046 & -77.3 & 0.796 & 0.384 & -60.8 \\            
 2010/Dec/04 & C1 & 1.120 & 0.035 & -85.0 & 0.695 & 0.515 & -59.5 \\
 2010/Dec/16 & C1 & 1.205 & 0.031 & -73.9 & 0.716 & 0.694 & -52.0 \\ \hline
  \end{tabular}
 \end{center} 
\end{table*}

\begin{table*}
 \caption{Summary of Gaussian model parameters of PKS~1510-089.} \label{tab:summary-of- PKS1510-091-parameters}
 \begin{center}
   \begin{tabular}{cccccccc} \hline\hline
 Epoch &  Component & Flux [Jy] & Radius [mas] & $\theta$ [deg] &  Major [mas] & Axial ratio &  $\phi$ [deg]     \\  \hline
 2010/Nov/18 & C1 & 1.968 & 0.008 & -88.2 & 0.538 & 0.349 & -33.6 \\
 &  C2 & 0.129 & 2.523 & -30.8 & 1.614 & 0.292 & -17.3 \\    
 2010/Nov/29 & C1 & 1.817 & 0.014 & -47.1 & 0.618 & 0.368 & -30.7 \\ 
 & C2 & 0.112 & 2.474 & -36.1 & 0.777 & 0.546 & -25.6 \\  
2010/Dec/04 & C1 & 1.771 & 0.022 & -50.7 & 0.544 & 0.436 & -30.4 \\
 & C2 & 0.081 & 2.118 & -33.6 & 0.000 & 1.000 & 0.0 \\ \hline
\end{tabular}
 \end{center} 
\end{table*}

\begin{table*}
 \caption{Summary of Gaussian model parameters of DA~406.} \label{tab:summary-of- DA406-parameters}
 \begin{center}
   \begin{tabular}{cccccccc} \hline\hline
 Epoch &  Component & Flux [Jy] & Radius [mas] & $\theta$ [deg] &  Major [mas] & Axial ratio &  $\phi$ [deg]     \\  \hline 
2010/Dec/21 & C1 & 1.256 & 0.007 & 7.647 & 1.491E-08 & 1.000  & -3.5 \\ 
 & C2 & 0.110 & 4.355 & 157.1 & 0.826 & 1.000 & 8.5 \\
 & C3 & 0.041 & 1.043 & 170.5 & 0.607 & 1.000 & -12.2 \\
2011/Jan/09+Jan/15 & C1 & 1.113 & 0.034 & -2.8 & 1.761e-07 & 1.000 &  -8.5 \\
 & C2 & 0.049 & 4.303 & 158.8 &  0.194 & 1.000 & 0.0 \\
 & C3 &  0.149 &  0.526 & 177.7 & 0.533 & 1.000 & -8.5 \\
2011/Feb/02 & C1 & 1.281 & 0.012 & -2.1 & 0.130 & 1.000 &  -36.2 \\
& C2  &  0.056 & 4.202 & 160.1 &  0.660 &  1.000 & -10.1 \\
 & C3  & 0.071 & 0.798 & 168.8 & 5.067e-07 & 1.000 & -5.6 \\ \hline
  \end{tabular}
 \end{center} 
\end{table*}

\begin{table*}
 \caption{Summary of Gaussian model parameters of NRAO~530.} \label{tab:summary-of- NRAO530-parameters}
 \begin{center}
   \begin{tabular}{cccccccc} \hline\hline
 Epoch &  Component & Flux [Jy] & Radius [mas] & $\theta$ [deg] &  Major [mas] & Axial ratio &  $\phi$ [deg]     \\  \hline 
2010/Nov/11 & C1 & 2.733  & 0.118  & -170.1  & 0.000  & 1.000  & 0.0  \\
 & C2 & 1.405  & 0.254  & 6.6  & 0.000  & 1.000  & 0.0  \\
 & C3 & 0.118  & 2.682  & -3.9  & 1.723  & 0.000  & -36.2  \\
2010/Nov/14 & C1 & 2.594  & 0.129  & -177.1  & 0.000  & 1.000  & 0.0  \\
 & C2 & 1.573  & 0.228  & 2.7  & 0.000  & 1.000  & 0.0  \\
 & C3 & 0.174  & 2.715  & -3.8  & 1.517  & 0.516  & -30.886  \\
2010/Nov/16 & C1 & 2.925  & 0.092  & -168.0  & 0.000  & 1.000  & 0.0  \\
 & C2 & 1.142  & 0.262  & 11.9  & 0.000  & 1.000  & 0.0  \\
 & C3 & 0.130  & 2.639  & -6.1  & 1.869  & 0.456  & -46.8  \\ \hline
  \end{tabular}
 \end{center} 
\end{table*}

\begin{table*}
 \caption{Summary of Gaussian model parameters of BL~Lac.} \label{tab:summary-of- BLLAC-parameters}
 \begin{center}
   \begin{tabular}{cccccccc} \hline\hline
 Epoch &  Component & Flux [Jy] & Radius [mas] & $\theta$ [deg] &  Major [mas] & Axial ratio &  $\phi$ [deg]     \\  \hline 
2011/Feb/08 & C1 & 2.477 & 0.054 & 171.7 & 0.598 & 0.000 & 41.5 \\
& C2 & 0.593 & 2.125 & -165.5 & 0.788 & 0.743 & -6.9 \\
2011/Feb/12 & C1 & 2.175 & 0.006 & 171.6 & 0.330 & 0.000 & 45.4 \\
& C2 & 0.843 & 2.038 & -167.8 & 2.137 & 0.306 & 10.3 \\
2011/Feb/13 & C1 & 2.489 & 0.015 & 158.1 & 0.423 & 0.333 & 20.0 \\
& C2 & 0.680 & 1.945 & -169.0 & 1.265 & 0.445 & 17.2 \\ \hline
  \end{tabular}
 \end{center} 
\end{table*}

\begin{table*}
 \caption{Summary of Gaussian model parameters of 3C~454.3.} \label{tab:summary-of-3C454.3-parameters}
 \begin{center}
   \begin{tabular}{cccccccc} \hline\hline
 Epoch &  Component & Flux [Jy] & Radius [mas] & $\theta$ [deg] &  Major [mas] & Axial ratio &  $\phi$ [deg]     \\  \hline 
2010/Nov/10 & C1 & 21.039 & 0.053 & 79.1 & 0 & 1 & 0 \\
& C2 & 3.987  & 0.363 & -102.7 & 0 & 1 & 0 \\     
2010/Nov/12 & C1 & 21.230 & 0.041 & 80.9 & 0 & 1 & 0 \\
& C2 & 2.771 & 0.423 & -100.2 & 0 & 1 & 0 \\ 
2010/Dec/11 & C1 & 21.487 & 0.046 & 76.0 & 0 & 1 & 0 \\
& C2 & 3.456  & 0.390 & -104.3 & 0 & 1 & 0 \\ \hline
  \end{tabular}
 \end{center} 
\end{table*}
 
\begin{figure*}[]
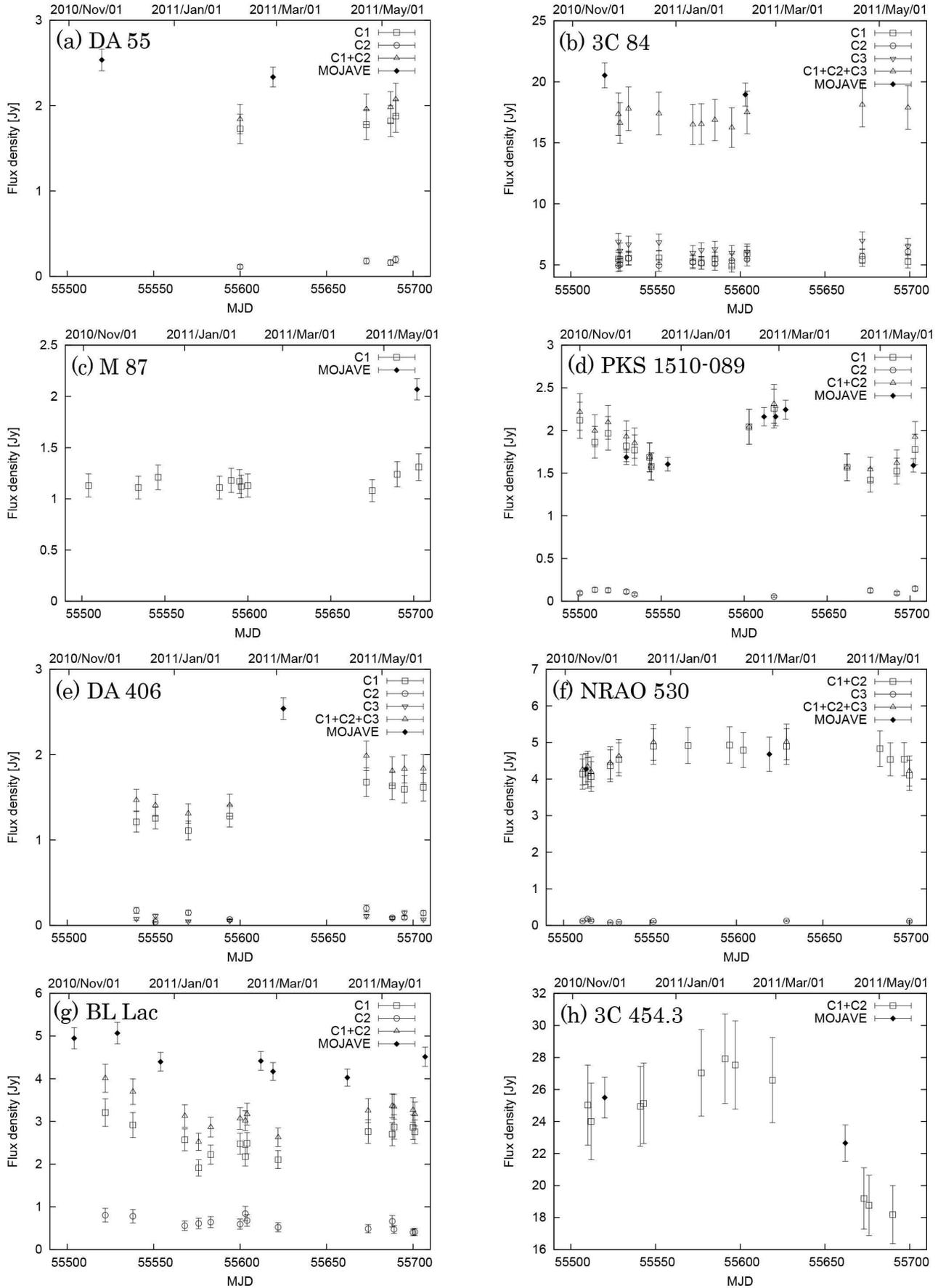
 
\begin{tabular}{cc}
\begin{minipage}{0.5\hsize}
\FigureFile(80mm,80mm){Figure11a.eps}
\end{minipage}
\begin{minipage}{0.5\hsize}
\FigureFile(80mm,50mm){Figure11b.eps}
\end{minipage} \\
\begin{minipage}{0.5\hsize}
\FigureFile(80mm,80mm){Figure11c.eps}
\end{minipage}
\begin{minipage}{0.5\hsize}
\FigureFile(80mm,80mm){Figure11d.eps}
\end{minipage}\\
\begin{minipage}{0.5\hsize}
\FigureFile(80mm,80mm){Figure11e.eps}
\end{minipage}
\begin{minipage}{0.5\hsize}
\FigureFile(80mm,80mm){Figure11f.eps}
\end{minipage}\\
\begin{minipage}{0.5\hsize}
\FigureFile(80mm,80mm){Figure11g.eps}
\end{minipage}
\begin{minipage}{0.5\hsize}
\FigureFile(80mm,80mm){Figure11h.eps}
\end{minipage}
\end{tabular}
\caption{The first 6-month light curve of (a) DA~55, (b) 3C~84, (c) M~87, (d) PKS~1510-089, (e) DA~406, (f) NRAO~530, (g) BL~Lac, and (h) 3C~454.3.  For the comparison, MOJAVE flux data is also plotted.}
\label{fig:[lightcurve]}
\end{figure*}

\begin{figure}[H]
\begin{center}
\includegraphics[width=7cm]{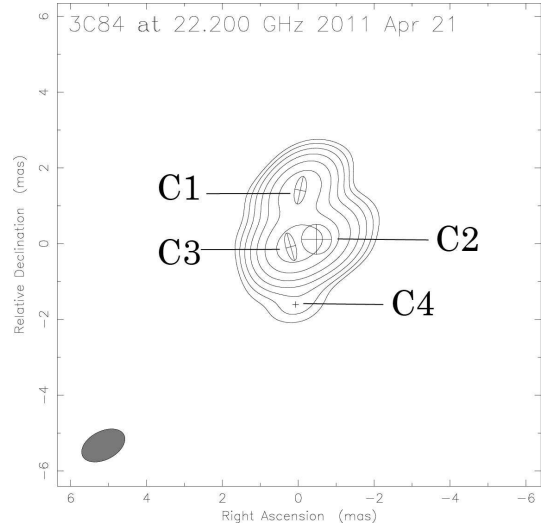}
\end{center}
\caption{Modelfit image of 3C~84 on 2011 April 21.  A point source C4 can be fitted at $\sim2$-mas south from the C3.}
\label{fig:[3C84_2011April21]}
\end{figure}

\begin{figure}[H]
\begin{center}
\includegraphics[width=7cm]{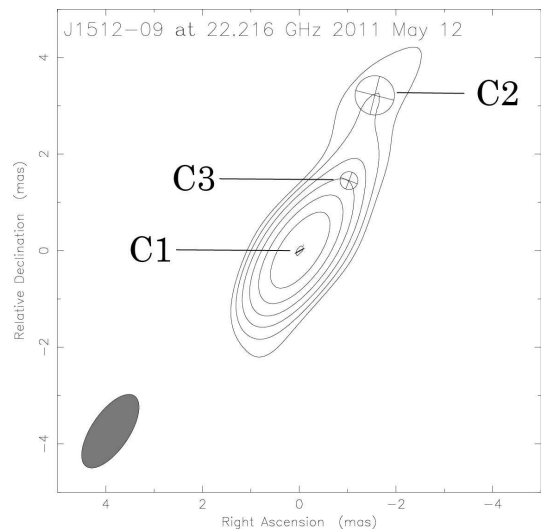}
\end{center}
\caption{Modelfit image of PKS1510-089 on 2011 May 12.}
\label{fig:[PKS1510_2011Nov12]}
\end{figure}

\end{document}